\documentclass[aps,prd,twocolumn,showpacs,superscriptaddress,nofootinbib,nobibnotes,longbibliography]{revtex4-1}  
\usepackage[latin1]{inputenc}
\usepackage{amsmath,amssymb} 
\usepackage{amsfonts} 
\usepackage{amsthm} 
\usepackage{latexsym}
\usepackage{graphicx} 
\usepackage{epstopdf} 
\usepackage[footnotesize]{caption} 
\usepackage{subfig}
\usepackage[colorlinks]{hyperref} 

\newcommand{\note}[1]{\text{\tiny{#1}}}

\newcommand{\dd}{\mathrm{d}}

\renewcommand{\rho}{\varrho}
\renewcommand{\epsilon}{\varepsilon}

\newcommand{\aaee}{\text{\ae}}
\newcommand{\AAEE}{\text{\AE}}
\def\beq{\begin{equation}}
\def\eeq{\end{equation}}
\def\bea{\begin{eqnarray}}
\def\eea{\end{eqnarray}}

\theoremstyle{plain}

\begin{document}
\title{Black hole horizons at the extremal limit in Lorentz-violating gravity}
\author{Nicola Franchini}
\affiliation{School of Mathematical Sciences, University of Nottingham, University Park, Nottingham, NG7 2RD, UK}
\author{Mehdi Saravani}
\affiliation{School of Mathematical Sciences, University of Nottingham, University Park, Nottingham, NG7 2RD, UK}
\author{Thomas P. Sotiriou}
\affiliation{School of Mathematical Sciences, University of Nottingham, University Park, Nottingham, NG7 2RD, UK}
\affiliation{School of Physics and Astronomy, University of Nottingham, University Park, Nottingham, NG7 2RD, UK}

\begin{abstract}
Lorentz-violating gravity theories with a preferred foliation can have instantaneous propagation. Nonetheless, it has been shown that black holes can still exist in such theories and the relevant notion of an event horizon has been dubbed ``universal horizon''. In stationary spacetimes the universal horizon has to reside in a region of spacetime where the Killing vector associated with stationarity is spacelike. This raises the question of what happens to the universal horizon in the extremal limit, where no such region exists anymore. We use a decoupling limit approximation to study this problem. Our results suggest that at the extremal limit, the extremal Killing horizon appears to play the role of a degenerate universal horizon, despite being a null and not a spacelike surface, and hence not a leaf of the preferred foliation.
\end{abstract}

\maketitle

\section{Introduction}

From a perspective of causality, there seem to be at least two distinct ways one can violate boost invariance --- an essential part of Lorentz symmetry. The first way is to introduce a preferred frame and have different types of excitations propagate at different speeds in that preferred frame, perhaps some of them superluminal and some of them subluminal.  The second way is to introduce a preferred foliation. This allows for the following: (i) a more general modification of the dispersion relation that can include higher-order terms of the type $\omega^2\propto k^2 +\alpha \,k^4/M_\star^2+\ldots$, where $\omega$ is the frequency, $k$ is the wave number, $M_\star$ is a characteristic mass scale associated with Lorentz symmetry breaking, and $\alpha$ is a dimensionless parameter; (ii) degrees of freedom that carry no time derivatives at all \cite{Donnelly:2011df,Blas:2011ni,Bhattacharyya:2015uxt}. Either way, one can have infinitely fast propagation (either for very large momenta or for all momenta respectively).

Superluminal propagation at finite or infinite speed requires one to rethink the concept of a black hole. The existence of black holes in the Universe might appear to be the ultimate vindication of the causal structure of general relativity and of Lorentz symmetry. However, before making such a statement, one actually needs to understand whether and how the concept of black holes fits into Lorentz-violating gravity theories. In fact, this is in its own right a very strong motive for studying Lorentz-violating theories in the first place.

In a Lorentz-violating theory in which excitations propagate with different but finite speeds in a preferred frame, it is rather straightforward to suitably generalise the definition of a black hole. Such excitations can be thought of as propagating along null cones of different effective metrics \cite{Eling:2006ec}. A stationary black hole spacetime will possess a succession of horizons, one for each excitation that travels at a unique speed. All of these horizons will be null surfaces of the corresponding metric. They will all cloak the singularity and the innermost one, the one corresponding to the fastest mode, will act as the boundary of the region that is causally disconnected from the exterior \cite{Eling:2006ec,Barausse:2011pu}. This scenario is well studied in the context of Einstein-aether theory ({\ae}-theory), which is the most general theory of a unit timelike vector (aether) coupled to Einstein gravity, containing only two derivatives of the unit vector $u^\mu$ \cite{Jacobson:2000xp,Jacobson:2008aj}.
Because of the unit constraint
\begin{equation}\label{eq:4velocity}
g_{\mu\nu}u^\mu u^\nu=1,
\end{equation}
the aether field $u^\mu$ never vanishes but instead always defines a preferred frame. The theory propagates spin-2, spin-1 and spin-0 modes that propagate along null trajectories of the effective metrics $g_{\mu\nu}^{(i)}\equiv g_{\mu\nu}- (s_{(i)}^2 -1) u_\mu u_\nu$, where $s_{(i)}$ is the speed of the corresponding spin-$i$ mode.

The issue of the existence and definition of black holes is more subtle in the case of a theory with a preferred foliation, since infinitely fast propagation seems particularly hard to reconcile with our conventional understanding of horizons. Additionally, notions like null trajectories, null surfaces, and null infinity become irrelevant for causality, which is now defined by the leaves of the foliation that correspond to constant preferred time surfaces (see Ref.~\cite{Bhattacharyya:2015gwa} for a more detailed discussion). Remarkably, a suitable notion of black hole does exist in such theories and it is  related with the behaviour of the foliation instead of the metric. In a generic asymptotically flat spacetime, the leaves of the foliation extend to spacelike infinity. If a leaf fails to satisfy this requirement it actually bounds a region that cannot communicate with infinity, since any signal can only penetrate a leaf in a single direction while traveling into the future. Hence, such a leaf defines a black hole (see Ref.~\cite{Bhattacharyya:2015gwa} for a rigorous definition) and it is  dubbed universal horizon \cite{Barausse:2011pu,Blas:2011ni}.

The concept of a universal horizon was first uncovered in the context of static, spherically symmetric solutions in Einstein-aether theory and the infrared limit of Ho\v rava gravity \cite{Barausse:2011pu,Blas:2011ni}. Ho\v rava gravity \cite{Horava:2009uw,Blas:2009qj,Sotiriou:2010wn} is actually a quantum gravity candidate. Its infrared limit, on which we will focus here, bears similarities with Einstein-aether theory, as will be discussed in more detail below. A key difference though is that unlike Einstein-aether theory, Ho\v rava gravity has a preferred foliation. That is, the equations are second order in time derivatives only in a specific foliation, which is thus causally preferred \cite{Jacobson:2010mx,Donnelly:2011df,Bhattacharyya:2015gwa}. Moreover, the theory exhibits instantaneous propagation even in the infrared limit \cite{Blas:2011ni,Bhattacharyya:2015uxt}. Hence, Ho\v rava gravity is the ideal test bed for a quantitative study of the concept of a universal horizon.

A foliation by spacelike hypersurfaces corresponds to the level surfaces of a scalar field whose gradient is everywhere timelike. Let us call this field $T$. The unit one-form
\begin{equation}\label{eq:HL4velocity}
u_\mu=\frac{\partial_\mu T}{\sqrt{g^{\mu\nu}\partial_\mu T\partial_\nu T}}.
\end{equation}
allows one to describe the foliation in a coordinate-invariant fashion and without the need to label the surfaces, as $u_\mu$ is invariant under reparametrizations of the type $T\to f(T)$. The reason that we chose $u_\mu$ for both the normalised gradient of $T$ here and the aether in Einstein-aether theory will become clear shortly. In stationary spacetimes it has been shown that one can  characterise the universal horizon by the local condition $u_\mu \chi^\mu=0$ provided that $a_\mu \chi^\mu \neq 0$, where $\chi$ is the Killing vector associated with stationarity and $a_\mu \equiv u^\nu \nabla_\nu u_\mu$ is the acceleration of $u_\mu$ \cite{Bhattacharyya:2015gwa}. The condition $u_\mu \chi^\mu=0$ bears a strong similarity with the local characterisation of event horizons as Killing horizons in general relativity and the condition $\chi^2=0$. The $a_\mu \chi^\mu \neq 0$ requirement came up as a technical restriction in the proof given in Ref.~\cite{Bhattacharyya:2015gwa}, but it appears to have a physical interpretation as a non-degeneracy condition for the universal horizon. In particular, it can be shown that $a_\mu \chi^\mu$ is constant along the universal horizon \cite{Bhattacharyya:2015gwa} and it has been argued to play the role of surface gravity \cite{Berglund:2012fk,Cropp:2013sea}.

Since $u_\mu$ is timelike per definition, the condition $u_\mu \chi^\mu=0$ can only be satisfied in a region of spacetime where $\chi^\mu$ is spacelike. Indeed, in all of the known stationary solutions the universal horizon lies in such a region \cite{Barausse:2011pu,Blas:2011ni,Berglund:2012bu,Barausse:2012ny,Barausse:2012qh,Sotiriou:2014gna,Barausse:2015frm,Saravani:2013kva,Meiers:2015rzm}. This raises the question of what happens to the universal horizon in the limit where a rotating or a charged black hole becomes extremal. In GR, in the extremal limit the Killing vector is everywhere timelike except on the (degenerate) Killing horizon, where it is null. Hence, it appears that no such solution can have a universal horizon. If instead one considers a non-extremal solution with a universal horizon, then the latter is expected to lie between two Killing horizons \cite{Bhattacharyya:2015gwa}. As one moves closer to extremality, the two Killing horizons should get closer to each other and to the universal horizon and there is no reason to expect that any singularity will appear even for an arbitrarily small deviation from extremality.

Motivated by the above, we study here the behaviour of the foliation in extremal and nearly extremal black holes in the infrared limit of Ho\v rava gravity. To be able to restrict ourselves to spherical symmetry that simplifies the analysis, we consider charged instead of rotating black holes. As we will discuss below, the assumption of spherical symmetry has the added advantage that our analysis will apply to Einstein-aether theory as well \cite{Jacobson:2010mx,Barausse:2012ny}. An additional restriction we impose is that we use the decoupling limit: i.e. we consider only the behaviour of the foliation in a fixed black hole background that solves Einstein's equations and ignore the backreaction of the foliation-defining field on the geometry. Though this might seem drastic, we believe that it provides technical simplification without compromising the qualitative understanding of the problem at hand.

\section{Einstein-aether theory and infrared limit of Ho\v{r}ava gravity}

Below we will use the infrared limit of  Ho\v{r}ava gravity. For the purposes of our analysis it is instructive to introduce the corresponding action and field equations through their relation to \ae-theory. The action of {\ae}-theory reads
\begin{equation}\label{eq:action-ae}
S=\frac{1}{16\pi G_\aaee}\int\dd^4x\sqrt{-g}(-R-M^{\alpha\beta}_{\ \ \mu\nu}\nabla_\alpha u^\mu\nabla_\beta u^\nu),
\end{equation}
where $R$ is the Ricci scalar and
\begin{equation}\label{eq:Maether}
M^{\alpha\beta}_{\ \ \mu\nu}=c_1 g^{\alpha\beta}g_{\mu\nu}+c_2\delta^\alpha_{\ \mu}\delta^\beta_{\ \nu}+c_3\delta^\beta_{\ \mu}\delta^\alpha_{\ \nu}+ c_4u^\alpha u^\beta g_{\mu\nu},
\end{equation}
with $c_i$ being dimensionless coupling constants whereas $G_\aaee$ is related to Newton constant $G_N$ as measured in a Cavendish experiment by $G_N=G_\aaee/(1-c_{14}/2)$ and we are following the convention  $c_{ij}=c_i+c_j$. The aether satisfies the unit timelike constraint \eqref{eq:4velocity} that can be imposed explicitly in \eqref{eq:action-ae} as a Lagrange multiplier term $\zeta(g_{\mu\nu}u^\mu u^\nu-1)$.

The equations of motion for the metric and aether fields can be obtained by varying the action with respect to $g^{\mu\nu}$ and $u_\mu$ and using the unit constraint to eliminate the Lagrange multiplier. This process yields
\begin{align}
G_{\mu\nu}-T^\aaee_{\mu\nu}& =0, \label{eq:Einstein}\\
\AAEE^\mu & =0, \label{eq:aeequation}
\end{align}
where $G_{\mu\nu}\equiv R_{\mu\nu}-\frac{1}{2}Rg_{\mu\nu}$ is the Einstein tensor,
\begin{equation}
\begin{aligned}\label{eq:aeSEtensor}
T^\aaee_{\alpha\beta}= & \nabla_\mu\left(J_{\left(\alpha\right.}^{\ \mu} u_{\left.\beta\right)}-J^\mu_{\ \left(\alpha\right.}u_{\left.\beta\right)}- J_{\left(\alpha\beta\right)}u^\mu\right) \\
 & +c_1\left[\left(\nabla_\mu u_\alpha\right)\left(\nabla^\mu u_\beta\right)-\left(\nabla_\alpha u_\mu\right)\left(\nabla_\beta u^\mu\right)\right] \\
 & +\left[u_\nu\left(\nabla_\mu J^{\mu\nu}\right)-c_4\dot{u}^2\right]u_\alpha u_\beta+c_4\dot{u}_\alpha\dot{u}_\beta-\frac{1}{2}L_\aaee g_{\alpha\beta},
\end{aligned}
\end{equation}
is the aether stress energy tensor,
\[
J^\alpha_{\ \mu} =M^{\alpha\beta}_{\ \ \mu\nu}\nabla_\beta u^\nu, \qquad \dot{u}_\nu=u^\mu\nabla_\mu u_\nu,
\]
and
\begin{equation}\label{eq:aethertensor}
\AAEE^\mu=\nabla_\alpha J^{\alpha\mu}-c_4\dot{u}_\alpha\nabla^\mu u^\alpha.
\end{equation}
As discussed in the Introduction, the unit constraint \eqref{eq:4velocity} implies that the aether cannot vanish in any spacetime, including the Minkowski spacetime. Instead, the aether defines a preferred threading by timelike trajectories and hence it introduces a preferred frame.

One could further restrict the aether to be hypersurface orthogonal \cite{Jacobson:2010mx}. Taking into account the unit constraint as well, hypersurface orthogonality can be imposed by the local condition \eqref{eq:HL4velocity}. The fact that $u_\mu$ is timelike implies that $T$ will always have a timelike gradient throughout spacetime. That is, $T$ defines a foliation by spacelike hypersurfaces and $u_\mu$ is the normal to the leaves of this foliation.

For the rest of this paper we will focus on the theory described by action \eqref{eq:action-ae} together with the condition \eqref{eq:HL4velocity}. This condition is imposed {\em before} the variation and $T$ is now the fundamental field. One can show (see Ref.~\cite{Barausse:2012qh} for a detailed discussion) that the equation of motion for $T$ is
\begin{equation}\label{eq:aetherHL}
\nabla_\mu\left(\frac{\AAEE^\mu}{\sqrt{\nabla^\nu T\nabla_\nu T}}\right)=0,
\end{equation}
The field equation that one obtains when varying with respect to the metric is given by Eq.~\eqref{eq:Einstein} with $u_\mu$ satisfying Eq.~\eqref{eq:HL4velocity}.  If one decides to use $T$ as a time coordinate and write the theory in the foliation defined by $T$, then the action coincides with the action of Ho\v rava gravity, truncated to two derivatives in that foliation \cite{Jacobson:2010mx}. Hence, we refer to the theory as the infrared limit of Horava gravity and the action \eqref{eq:action-ae} supplemented by the condition \eqref{eq:HL4velocity} is its manifestly covariant formulation.

In its full glory, Ho\v rava gravity includes terms that contain higher order spatial derivatives. These terms lead to higher-order dispersion relations and infinitely fast propagation, as discussed in the Introduction. At the same time they improve the UV behaviour of the theory \cite{Horava:2009uw}. In principle one needs to include in the action all such terms that have up to six spatial derivatives and are compatible with the symmetries of the theory, {\em i.e.} diffeomorphisms that respect the foliation, $t\to \tilde{t}(t)$ and $x^i\to \tilde{x}^i(x^i,t)$. There is a large number of such terms that makes the theory less tractable and raises concerns about predictability. As a result, various restricted versions that consistently reduce the number of UV terms have appeared \cite{Sotiriou:2009gy,Weinfurtner:2010hz,Horava:2010zj,Vernieri:2011aa,Vernieri:2012ms} but this reduction usually comes at the cost of infrared viability \cite{Sotiriou:2009bx,Charmousis:2009tc,Blas:2009yd,Koyama:2009hc,Papazoglou:2009fj,Blas:2009ck,Kimpton:2010xi,Mukohyama:2010xz}.

Here we will not be concerned with the UV completion of Ho\v rava gravity. Hence, we will ignore the higher order terms and focus on the infrared limit as defined above.
Already in that limit, the theory has the characteristics that we want: a preferred foliation and instantaneous propagation. This last statement might not seem obvious, because the infrared limit does away with the higher order terms in the dispersion relations and renders them linear.  Some intuition can be gained by inspecting Eq.~\eqref{eq:aetherHL}. This equation is fourth order in derivatives of $T$, as $\AAEE_\mu$ contain two derivatives of $u_\mu$, $u_\mu$ contains a further derivative of $T$ and there is a extra explicit derivative. However, if one chooses $T$ as a time coordinate then the only nonvanishing component of $u_\mu$ is $u_T=N$, where $N$ is the lapse of the foliation defined by $T=const$ surfaces. Moreover, $\AAEE^\nu u_\nu=0$ identically: {\em i.e.},~$\AAEE^\nu$ is orthogonal to $u_\nu$ and it lies entirely on $T=const$ surfaces. Hence, the derivative that appears explicitly in Eq.~\eqref{eq:aetherHL} becomes a spatial divergence on a leaf. These suggest that the theory is second order in time derivatives only in the preferred foliation and that there is some elliptic equation that needs to be solved on each slice. Indeed, more detailed analysis has shown that both statements are true \cite{Donnelly:2011df,Blas:2011ni,Bhattacharyya:2015uxt}.

\section{Setup in the decoupling limit}

\subsection{The Reissner-Nordstr{\"o}m metric}

In general relativity, the Reissner-Nordstr{\"o}m (RN) metric describes a spherically symmetric object with electromagnetic charge. The parameters needed to describe this solution are its mass $M$ and electric charge $Q$. The metric reads:
\begin{equation}\label{eq:RNmetric}
\dd s^2=f(r)\dd t^2-\frac{\dd r^2}{f(r)}-r^2\dd\theta^2-r^2\sin^2\!\theta\,\dd\phi^2,
\end{equation}
where $f(r)=1-\frac{r_s}{r}+\frac{r_q^2}{r^2}$, $r_s=2G_N M$ and $r_q^2=Q^2 G_N/(4\pi\epsilon_0)$. The RN metric admits a Killing vector that is timelike at infinity and it is given in our coordinate as $\chi^\mu=(1,0,0,0)$. The condition for the Killing vector being null determines the position of the Killing horizons that are located at
\begin{equation}\label{eq:RNradius}
r_\pm=\frac{r_s\pm\sqrt{r_s^2-4r_q^2}}{2}.
\end{equation}
The cosmic censorship conjecture \cite{Penrose:1969pc} states that in order to avoid naked singularities $r_q\leq r_s/2$. Thus, the extremality condition occurs as $r_q=r_s/2$ for which the two radii of Eq. \eqref{eq:RNradius} coincide at $r=r_\note{ex}=r_s/2$.
We introduce the parameter $\epsilon$, useful to express the deviation from extremality
\[
r_q=\frac{r_s}{2}\sqrt{1-\epsilon^2}.
\]

\subsection{Determining the foliation at decoupling}

As stated in the Introduction, instead of considering the full field equations \eqref{eq:Einstein} and \eqref{eq:aetherHL}, we will work in the decoupling limit. That is, we will neglect entirely the backreaction that $T$ has on the metric and we will solve Eq.~\eqref{eq:aetherHL} on the RN black hole background that was reviewed above.
It is worth mentioning that  Eqs.~\eqref{eq:aetherHL} and \eqref{eq:aeequation} share all static, spherically symmetric solutions with flat asymptotics \cite{Barausse:2012ny}.
Hence, our solutions will also determine fully the configuration of the aether in \ae-theory in the decoupling limit.

Assuming that $u_\mu$ respects the Killing symmetries of the metric, the most generic form for the scalar field $T$ is
\begin{equation}\label{eq:scalaransatz}
T(t,r)=t+C(r)
\end{equation}
with $C(r)$ a generic function of $r$. The four-velocity vector will be
\begin{equation}\label{eq:4velansatz}
u_\mu=N(r)\partial_\mu T,
\end{equation}
and then, imposing condition \eqref{eq:4velocity} yields
\begin{equation}\label{eq:Ccond}
\frac{\partial C(r)}{\partial r}=\frac{\sqrt{N^2(r)-f(r)}}{f(r)N(r)}.
\end{equation}
Recall that $a_\mu\equiv \dot{u}_\mu$. With this ansatz one has
\begin{equation}\label{eq:acceleration}
a_\mu=-\left(\sqrt{N^2(r)-f(r)}\frac{\partial N(r)}{\partial r},\frac{N(r)}{f(r)}\frac{\partial N(r)}{\partial r},0,0\right).
\end{equation}
As already discussed above, for static, spherically symmetric solutions with flat asymptotics, Eqs.~\eqref{eq:aetherHL} and \eqref{eq:aeequation} yield the same solutions  \cite{Barausse:2012ny} and in our setup they can be reduced to the following equation
\begin{equation}\label{eq:Nsecond}
N''=\frac{N s^2\left(1-\epsilon^2-N'^2-2\frac{\tilde f(\xi)-N^2}{\xi^2}- \frac{(\tilde f'(\xi)-2N N')^2}{4(\tilde f(\xi)-N^2)}\right)}{\tilde f(\xi)+(s^2-1)N^2} ,
\end{equation}
where a prime denotes differentiation with respect to $\xi=r_s/2r$. $\tilde f(\xi)=f(\frac{r_s}{2\xi})=(1-\epsilon^2)\xi^2-2\xi+1$ is the $tt$ component of the metric and
\begin{equation}\label{eq:spin0speed}
s^2=\frac{c_{123}}{c_{14}}.
\end{equation}

\section{Foliating extremal and nearly extremal black holes}

\subsection{Extremal case ($\epsilon=0$)}\label{sec:extremal}
As we have already discussed in the Introduction, it appears to be impossible to satisfy the condition $u\cdot\chi=0$ in  an extremal RN spacetime. The Killing vector $\chi^\mu$ is everywhere timelike or null, but never becomes spacelike. $u^\nu$ is always timelike and the inner product of a timelike vector with a null or timelike vector $\chi^\mu$ is never zero.

Recall that $u\cdot\chi=0$ serves as a local characterisation of a universal horizon $r_\note{UH}$, provided that $a\cdot \chi \neq 0$, where $a_\mu$ is the acceleration of $u_\mu$. The condition $a\cdot \chi \neq 0$ has been suggested as a non-degeneracy condition, similar to requiring non-vanishing surface gravity for Killing horizons. Extremal horizons might well be degenerate (as in GR) and hence the arguments above regarding $u\cdot\chi$ are not enough to conclude that there will not be a universal horizon in the extremal limit. In fact, there is no known local characterisation of an extremal universal horizon, so it is not possible to check without having the full solution and the global causal structure.

There is actually a rather trivial solution to Eqs.~\eqref{eq:aeequation} and \eqref{eq:aetherHL} when $\epsilon=0$, $T(t,r)=t$. It is very tempting to dismiss this solution, due to the behaviour of the $T(t,r)=t$ foliation on the Killing horizon. At $r=r_\note{ex}=r_s/2$ a $t=const$ surface is null, so it cannot be a leaf of a regular foliation by spacelike hypersurfaces. Said otherwise, the normal vector to these surfaces is parallel to the Killing vector $\chi$, and (if normalized) it satisfies aether's equation of motion \eqref{eq:aeequation}, but at the same time it needs to be null at  $r=r_\note{ex}=r_s/2$. This implies that the foliation is not well defined there and cannot be smoothly continued past that surface.

Consider, however, $t=const$ surfaces in the exterior of the extremal RN spacetime, which corresponds to the analytic solution of Eq.~\eqref{eq:Nsecond},
\begin{equation}
\label{analytic}
N_0(\xi)=1-\xi.
\end{equation}
The foliation leaves are everywhere spacelike and they have the right asymptotic behaviour $N_0(0)=1$. Indeed, for $r\rightarrow\infty$, the aether $u^\mu$ aligns with the killing vector $\chi^\mu$ and this corresponds to the condition $N(0)=1$. Moreover, $N_0$ becomes zero at $\xi=1$ (corresponding to $r=r_\note{ex}$). This last features implies that  all leaves asymptote and pile up on the $r=r_\note{ex}=r_s/2$, which is exactly the behaviour they are expected to have as they approach a universal horizon. Hence, this seemingly unphysical, trivial solution is actually compatible with the naive expectation that the universal horizon and the Killing horizon coincide in the extremal limit. In the next section, we will demonstrate numerically that the exterior of the universal horizon in nearly extremal RN black holes does indeed approach this analytic solution as $\epsilon \to 0$.

\subsection{Nearly extremal case}

We now move on to nonextremal black holes. The plan is to generate the foliation numerically by integrating Eq.~\eqref{eq:Nsecond}. First it is important to understand how many independent parameters our solutions will have. Clearly, given that Eq.~\eqref{eq:Nsecond} is a second-order ODE, generically there will be a 2-parameter family of solutions for a black hole of given mass and electric charge. However, these solutions are expected to be singular whenever the denominator of the right-hand side of Eq.~\eqref{eq:Nsecond} vanishes. The locations of these singularities are determined by the roots of the following equation
\begin{equation}\label{eq:spin0hor}
(1-\epsilon^2)\xi^2-2\xi+1+(s^2-1)N(\xi)^2=0,
\end{equation}
which we collectively denote as $\xi_c$.

As discussed earlier, the spin-0 mode propagates along null geodesics of the effective metric, $g_{\mu\nu}^{(0)}\equiv g_{\mu\nu}- (s_0^2 -1) u_\mu u_\nu$. This can be explicitly seen by performing perturbative analysis around an arbitrary background for Eq.~\eqref{eq:Einstein} and either Eq.~\eqref{eq:aeequation} or Eq.~\eqref{eq:aetherHL}. The square of the speed of the spin-0 mode that is present in both \ae-theory and Ho\v rava gravity is
\begin{equation}
s_0^2=\frac{c_{123} (2-c_{14})}{c_{14}(1-c_{13}) (2+c_{13}+3 c_2)}.
\end{equation}
At the decoupling approximation that we are using here, backreaction is neglected and one solves Eq.~\eqref{eq:aeequation} or Eq.~\eqref{eq:aetherHL} in a fixed background. One can now determine the effective metric just by linearising Eq.~\eqref{eq:aeequation} or Eq.~\eqref{eq:aetherHL}. This yields $g_{\mu\nu}^{(0)}\equiv g_{\mu\nu}- (s^2 -1) u_\mu u_\nu$, where $s$ is as in Eq.~\eqref{eq:spin0speed}. That is, $s$ is the speed of the spin-0 mode at decoupling. Indeed, in the limit where all $c_i\to 0$, one has that $T^\aaee_{\mu\nu}\to 0$  and $s_0 \to s$, as expected.

Killing horizons of $g_{\mu\nu}^{(0)}$ are referred to as {\em spin-0 horizons} and act as causal boundaries for the spin-0 mode. Interestingly, their locations are also determined by roots of Eq.~\eqref{eq:spin0hor}, $\xi_c$, in our setup. In fact the denominator of the right-hand side of Eq.~\eqref{eq:Nsecond} is the norm of the Killing vector that generates the spin-0 horizon. This behaviour is expected, as it has been seen on all previous works (e.g.~\cite{Eling:2006ec,Barausse:2011pu,Blas:2011ni}).

This suggests that the most general solution will be singular on one or more spin-0 horizons.\footnote{Note that here we are interested in solutions with a universal horizon. The latter will always be cloaked by a spin-0 horizon \cite{Bhattacharyya:2015gwa}. In static, spherically symmetric solutions this follows easily from the fact that $u\cdot \chi=0$, which has to hold on the universal horizon, requires $\chi^\mu$ to be spacelike with respect to the spin-0 metric (or any spin-$i$ metric as defined above). Since $\chi^\mu$ is timelike at infinity, there needs to be a spin-0 horizon outside the universal horizon.}
There is always a 1-parameter subfamily of solutions that is regular on at least one spin-0 horizon of choice. To see this, recall that a solution to Eq.~\eqref{eq:Nsecond} can be generated starting from any given radius and selecting two pieces of ``initial data'', the values of $N$ and $N'$ there (that correspond to the two parameters of the general family).  One can then choose to start from a spin-0 horizon located at $\xi_c$ and denote
$N_c\equiv N(\xi_c)$. Assuming that $N_c\neq0$ and imposing that $N'(\xi_c)\equiv N'_c$ satisfies the bond
\begin{equation}\label{eq:NprimeS0}
N'_c=\frac{1-\xi_c(1-\epsilon^2)- s\sqrt{\epsilon^2+\frac{2N_c^4s^2(s^2-1)}{\xi_c^2}}}{N_c(s^2-1)},
\end{equation}
makes the numerator of the right-hand side of Eq.~\eqref{eq:Nsecond} vanish and one can integrate outwards and inwards and generate a solution that is regular on the spin-0 horizon and its vicinity. Given that the bond above relates the two pieces of initial data, there will be a 1-parameter family of such solutions.

For what comes next, we will apply this prescription to the outermost spin-0 horizon. The solutions of the 1-parameter family one obtains when integrating outwards toward infinity ($\xi\rightarrow0$) do not generically satisfy the asymptotic condition $N(0)=1$. Imposing this condition requires tuning the remaining piece of initial data. 

In practice, we impose the asymptotic condition via shooting. We impose that $N-1$ vanishes to a desired tolerance at a large but finite radius $r_\note{max}\equiv r_s/(2\xi_\note{min})$. We set the tolerance to be $\mathcal{O}(\xi_\note{min})$, as the general expectation is that $N$ will have a $1/r$ decay asymptotically ($\xi_\note{min}$ is chosen in the  range $[10^{-4},10^{-6}]$, depending on the solution).  Imposing the regularity condition \eqref{eq:NprimeS0} is not trivial when integrating numerically, because one needs to start the integration from a point where both the numerator and the denominator of the right- hand side of Eq.~\eqref{eq:Nsecond} vanishes. To avoid this problem, we expand $N(\xi)$ in a Taylor series in $\delta=\xi-\xi_c$ ($\delta$ can be positive or negative depending on the direction we are solving the equation) and we integrate  analytically from $\xi_c$ to $\xi_c+\delta$, up to fifth order, with $\delta=10^{-5}$. We then start the numerical integration from this point. The error of the perturbative solution at $\xi_c+\delta$ is then $\mathcal{O}(\delta^5)\sim10^{-25}$, comparable to the machine accuracy, i.e. 24 significant digits for our real numbers.

In general, it is conceivable that a boundary value problem gives rise to multiple solutions with the same boundary conditions. According to Eq. \eqref{eq:spin0hor}, $N_c$ can only take values in the interval $[0,\epsilon/\sqrt{(s^2-1)(1-\epsilon^2)}]$  for $s\ge1$. We have explored this interval numerically ($\sim$ 100 points) and we did not find another solution that satisfies the asymptotic condition $N(0)=1$. Though this is not a rigorous proof that our solution is unique,  considering numerical limitations the outcome is rather reassuring.

Starting with the same values of $N$ and $N'$ that gave a solution with the right asymptotic behaviour and regular on the outermost spin-0 horizon, one can then integrate inwards and generate the interior down to a desired radius. This process is quite straightforward up to (and including) the universal horizon, which is the location where $N$ vanishes. In the next section we present and discuss these solutions, i.e. the exterior of the universal horizon. If one attempts to continue this integration well past the universal horizon, a complication arises: eventually one will encounter a second spin-0 horizon.\footnote{In most cases that can be referred to as the inner spin-0 horizon, but in some cases there might be more than two spin-0 horizons in total.} The solution does not need to be smooth there and no further regularity condition can be imposed. So, they are essentially two options: (a) the solution is ``accidentally'' smooth; (b) one will encounter a singularity. In Sec. \ref{interior} we will present strong evidence that supports the latter case, but we will also argue that this does not affect the solution for the exterior of the universal horizon, even when  $\epsilon \to 0$.

\subsubsection{Exterior of the universal horizon}
\label{exterior}

There is degeneracy between changing the mass of the RN black hole and rescaling the radial coordinate in our equations. This implies that one can find a solution for a fixed mass and then obtain all other solutions by rescaling $r$. We have generated exterior solutions\footnote{From here onwards we will be referring to solutions that describe the exterior of the universal horizon simply as ``exterior solutions''.} for different values of $s$ but there is little qualitative difference between them. Hence, we will present here only solutions for $s=1.5$ and for different values of $\epsilon$.

\begin{figure}
\centering
\includegraphics[width=\columnwidth]{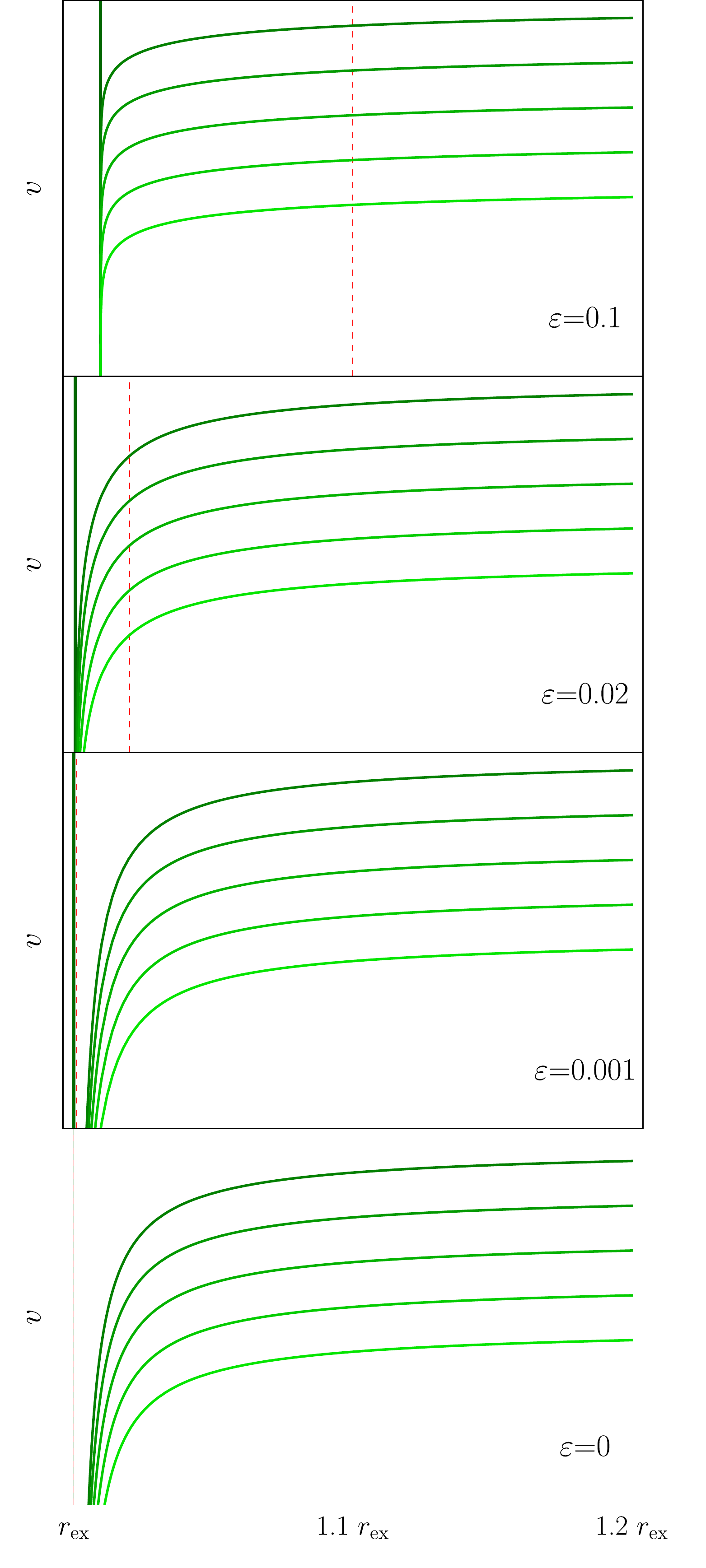}
\caption{Leaves of preferred foliation for different values of $\epsilon$, and $s=1.5$. The green vertical solid line shows the universal horizon and the red vertical dashed line shows the outer Killing horizon.
The bottom panel shows $T=t=$const in extremal RN, {\em i.e.}~the analytic solution in Eq.~\eqref{analytic}.
}\label{fig:Tconst}
\end{figure}

The $T=const$ surfaces (surfaces orthogonal to aether) correspond to leaves of the preferred foliation. $T$ is defined from Eq. \eqref{eq:scalaransatz} as $T=t+C(r)$. In outgoing Eddington-Finkelstein  coordinates $(v,r_*)$, where $v=t+r_*$ and $r_*$ is the  \textit{tortoise coordinate} defined by $\dd r_*=\dd r /f(r)$, one has
\begin{equation}\label{Tleaves}
T(v,r)=v+\int^r \left(\frac{\partial C}{\partial r'}-\frac{1}{f(r')}\right)\dd r'.
\end{equation}
In Fig. \ref{fig:Tconst} we show sets of foliation leaves for different values of $\epsilon$ in a $(r,v)$ plane. The leaves have a different behaviour depending on the value of $\epsilon$. As expected, as extremality is approached, the Killing horizon (red vertical dashed line) moves closer and closer to the universal horizon (green vertical solid line). In all non-extremal cases the leaves of the foliation asymptote to the universal horizon in the same fashion qualitatively. However, as one moves closer to extremality the foliation becomes denser, {\em i.e.}~one would need to cross more leaves to move radially from a fixed $r$ to another, smaller fixed $r$.

Another illuminating way to compare the foliations is to define a new spatial coordinate such that the locations of the Killing horizon of the metric and universal horizon $r_\note{UH}$ do not change for different values of $\epsilon$. Indeed, one can define such a coordinate as
\begin{equation}\label{eq:defw}
w=\frac{r-r_\note{UH}}{r-r_+},
\end{equation}
where  $w(r_\note{UH})=0$ and $w(r_+)=1$.

\begin{figure}
\centering
\includegraphics[width=\columnwidth]{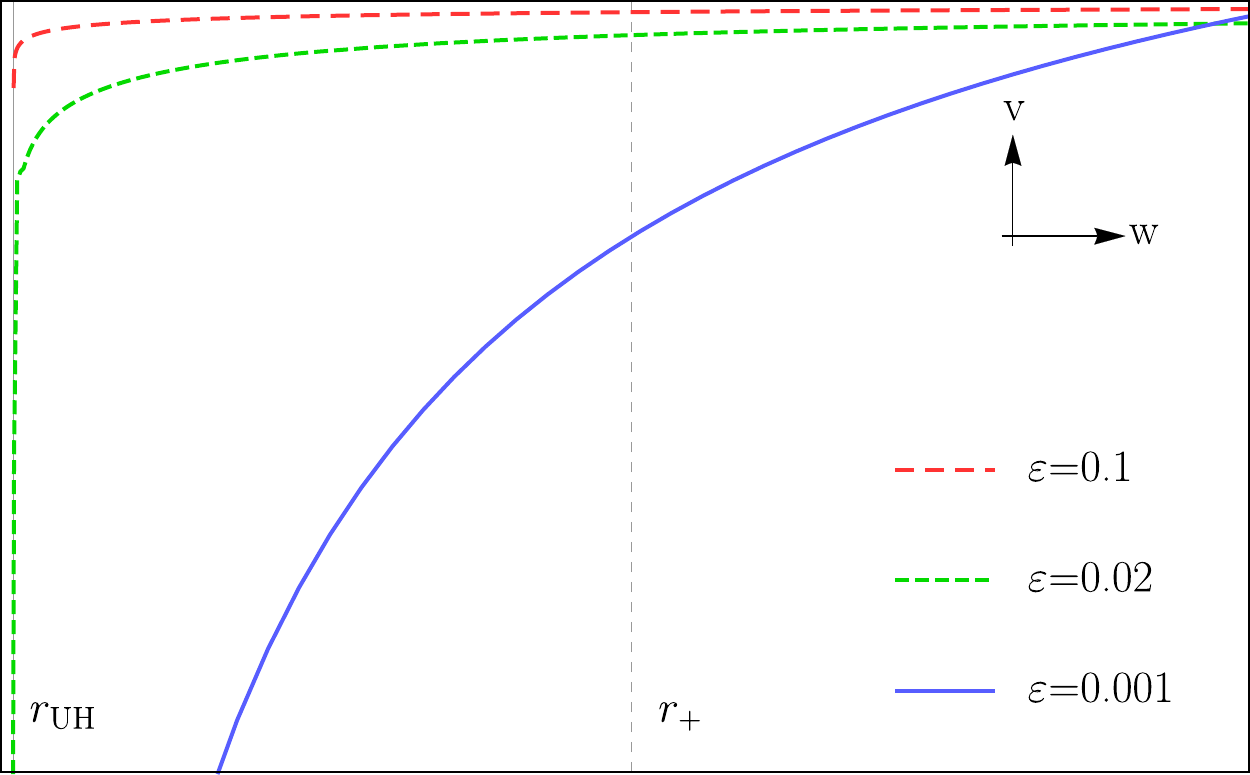}
\caption{One leaf of preferred foliation for different values of $\epsilon$, and $s=1.5$ in a $(w,v)$ plane, with $w$ defined in Eq.~\eqref{eq:defw}. The $w$ coordinate is chosen such that the positions of  $r_\note{UH}$ and $r_+$ are $\epsilon$-invariant.}\label{fig:leavesnorm}
\end{figure}

In Fig. \ref{fig:leavesnorm} we show one leaf from three different foliations ($\epsilon=0.1,~0.02,~0.001$) in a $(w,v)$ plane. The leaves cross at a point $(w_0,v_0)$ but we remark that, due to the $\epsilon$-dependent definition of $w$, $(w_0,v_0)$ does not correspond to a single point in $(r,v)$ coordinates. One can reproduce the entire foliation simply by shifting up and down the leaves we are showing in this plot.  The range of the plot has been chosen rather freely, approximately twice the distance between the Killing horizon  and the universal horizon. This plot clearly exhibits the different ways leaves approach the universal horizon as one moves closer to extremality.

The bottom panel of  Fig.~\ref{fig:Tconst}, corresponding to the extremal case and the analytic solution \eqref{analytic} , shows remarkable similarity with the panel right above it that corresponds to $\epsilon=0.001$. A remarkable difference is that the analytic solution breaks down on the extremal horizon and, hence, there is no foliation leaf that coincides with (or crosses) this location. In what follows we attempt to quantify the agreement between the numerical solutions and the analytic one as $\epsilon \to 0$.

\begin{figure}
    \centering
    \subfloat[][\label{fig:difflarge}]{\includegraphics[width=\columnwidth]{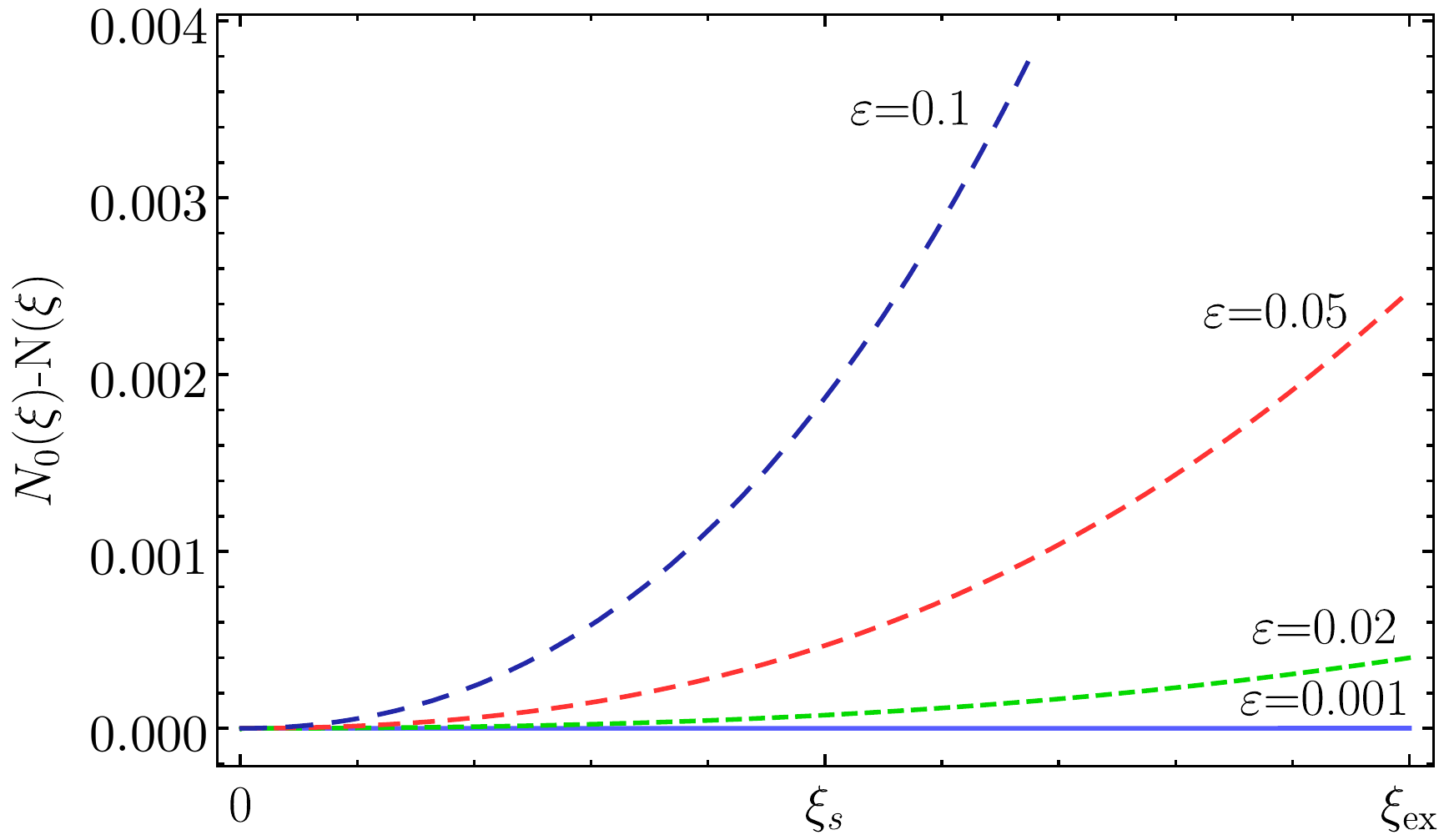}}
    \quad
    \subfloat[][ \label{fig:diff}]{\includegraphics[width=\columnwidth]{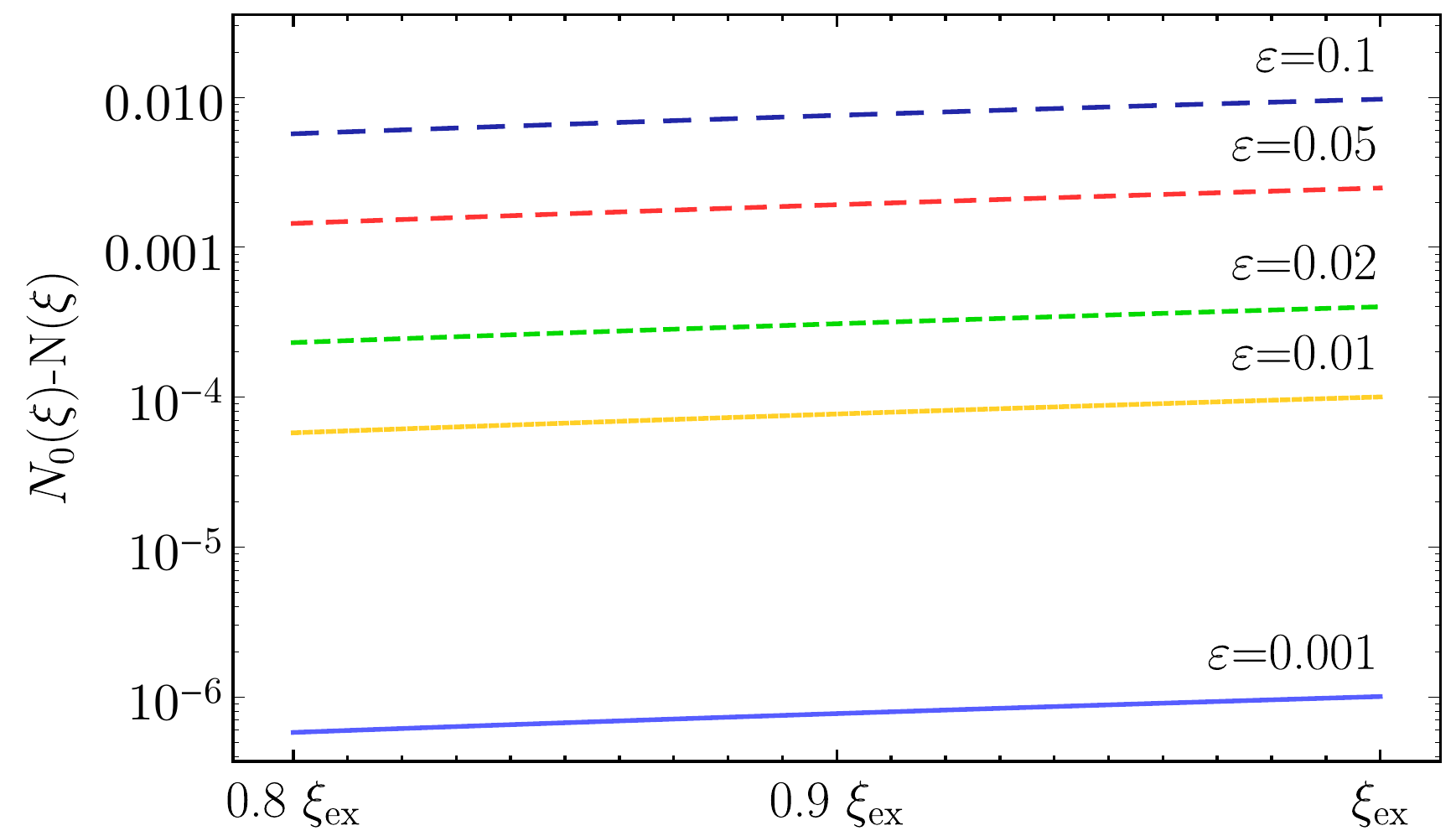}} \\
    \caption{The difference between the extremal, analytic solution in Eq. \eqref{eq:Nsecond} and numerical solutions $N_\epsilon(\xi)$ for different values of $\epsilon$ and $s=1.5$. In panel (a) one can see that all solutions approach $N_0$ at infinity. They deviate from it at smaller radii but the deviation decreases as one approaches extremality. Panel (b) is a logarithmic plot of $N_0(\xi)-N_\epsilon(\xi)$ in a region close to the radius of the extremal horizon, $\xi_\note{ex}$. $N_0(\xi)-N_\epsilon(\xi)$ scales as $\epsilon^2$. }
\end{figure}

Fig. \ref{fig:difflarge} shows the difference between the analytic, extremal solution $N_0(\xi)$ and the numerical solutions $N_\epsilon(\xi)$. $\xi_s$ and $\xi_\note{ex}$ are the corresponding values to $r_s$ and $r_\note{ex}$, respectively, {\em i.e.},~$\xi_s=0.5$ and $\xi_\note{ex}=1$. For all values of $\epsilon$, $N_0(\xi)$ serves as a good approximation at large radii, as expected. There is deviation as one approaches the black hole horizon, but it clearly decreases as $\epsilon$ decreases.  In Fig. \ref{fig:diff} one can clearly read off the scaling of these deviations. Indeed, this plot suggests that the numerical solutions differ from $N_0(\xi)$ by an $\epsilon^2$ correction.

\begin{figure}
  \centering
  \includegraphics[width=\columnwidth]{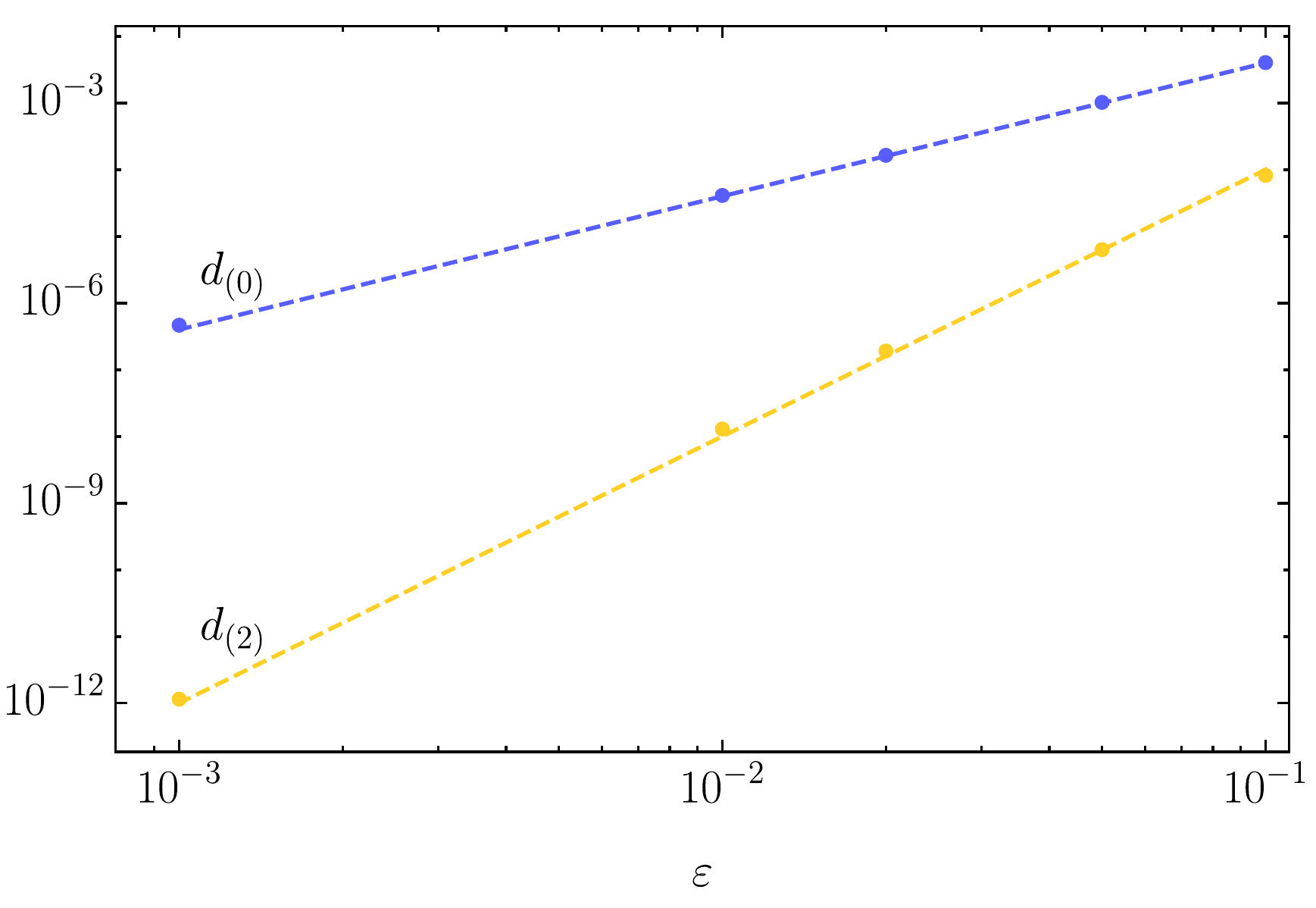}
  \caption{The blue dots show $d_{(0)}$ for different values of $\epsilon$ and $s=1.5$. The blue dashed line shows $\epsilon^2$ behaviour. The yellow dots show $d_{(2)}$ for different values of $\epsilon$ and $s=1.5$. The yellow dashed line shows $\epsilon^4$ behaviour.} \label{fig:deviation}
\end{figure}

To explicitly show this behaviour, we define the distance $d_{(0)}$ between the numerical solutions $N_\epsilon(\xi)$ and $N_0(\xi)$ (in the region $[0,\xi_\note{ex}]$) as follows
\beq\label{eq:d0}
d^2_{(0)}(\epsilon)=\int_0^{\xi_\note{ex}}\frac{\dd\xi}{\xi_\note{ex}}\Big(N_\epsilon(\xi)-N_0(\xi)\Big)^2.
\eeq
We compute $d_{(0)}$ for different values of $\epsilon$ and plot them in Fig.~\ref{fig:deviation}. In fact, upon further investigation one finds that the leading order term to $N_\epsilon(\xi)-N_0(\xi)$ is expected to be $-\frac{\epsilon^2}{2}\xi^2(1+\xi)$. To demonstrate this, we define and compute the distance
\beq\label{eq:d2}
d^2_{(2)}(\epsilon)=\int_0^{\xi_\note{ex}} \frac{\dd\xi}{\xi_\note{ex}}\left(N_\epsilon(\xi)-N_0(\xi)+\frac{\epsilon^2}{2}\xi^2(1+\xi)\right)^2,
\eeq
and investigate its behaviour for different values of $\epsilon$.  Fig. \ref{fig:deviation} shows $d_{(2)}$ as well, and it reveals that it scales as $\epsilon^4$. This confirms that  $N_\epsilon(\xi)-N_0(\xi) =-\epsilon^2\xi^2(1+\xi)/2+{\cal O}( \epsilon^4)$.

\begin{figure}
  \centering
  \includegraphics[width=\columnwidth]{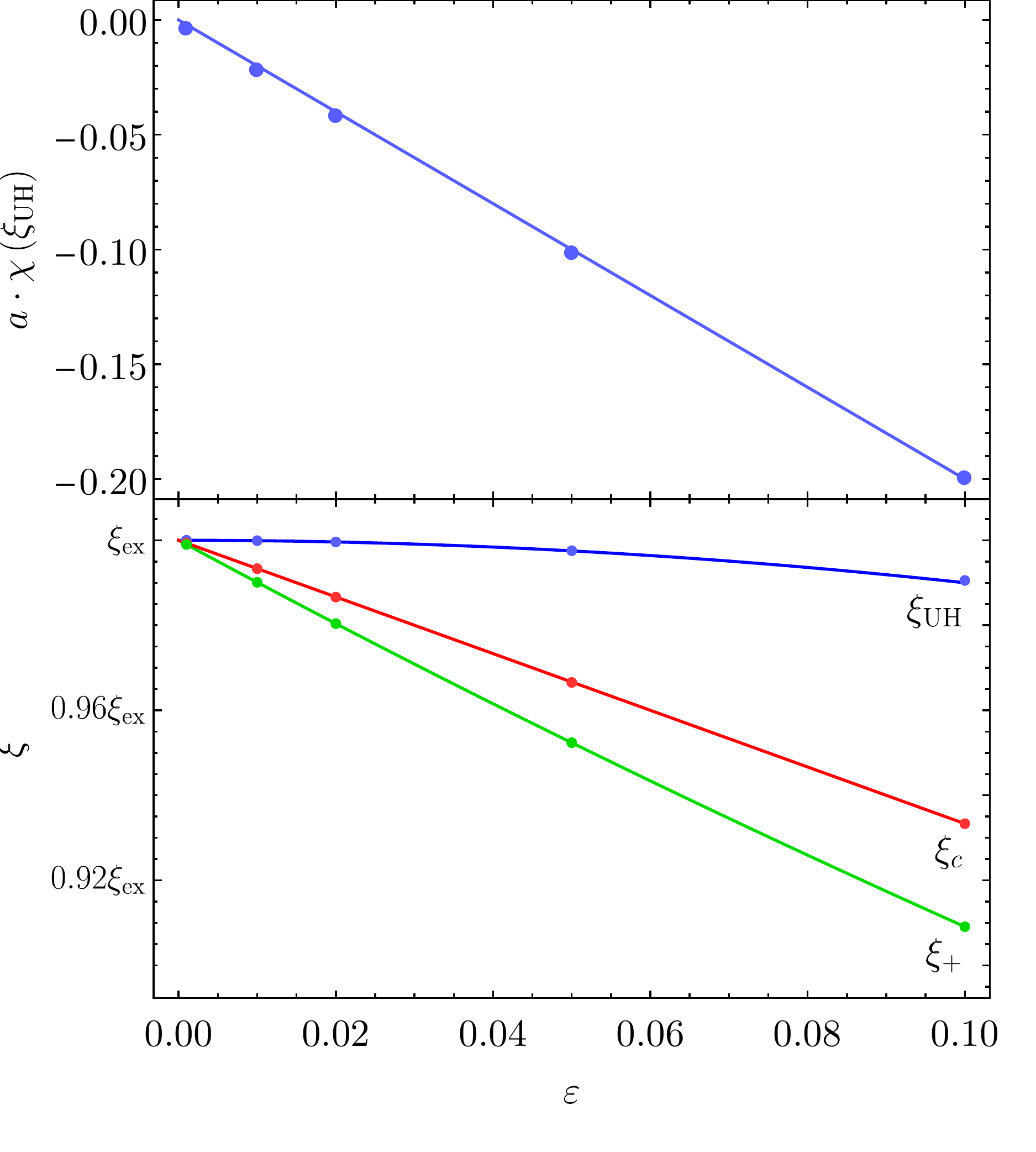}
  \caption{The upper panel shows  $a_\mu\chi^\mu$ evaluated at the universal horizon for various values of $\epsilon$ (blue dots). In this range there is a clear $-2\epsilon$ scaling (blue solid line). The lower panel shows the location of the universal horizon $\xi_\note{UH}$, outer spin-0 horizon $\xi_c^+$ and outer Killing horizon $\xi_+$ as a function of $\epsilon$. Dots are numerical solutions and solid lines are $\xi_H=1-\epsilon^2$, $\xi_c^+=1-\epsilon/s$ and $\xi_+=1-\epsilon$. In both plots the spin-0 sound speed is $s=1.5$.} \label{fig:epsi}
\end{figure}

In Fig. \ref{fig:epsi} we show some more features of the numerical solutions found. The upper panel shows $a\cdot \chi$ evaluated on the universal horizon. As one approaches extremality $a\cdot \chi\to 0$, in agreement with what is expected from the analytical solution $N_0$, for which $a\cdot \chi=0$ everywhere. Our approximate fit $N_\epsilon(\xi)\simeq 1-\xi-\epsilon^2 \xi^2(1+\xi)/2$ above yields $a\cdot \chi (\xi_\note{UH})\simeq-2\epsilon$, compatible with Fig. \ref{fig:epsi}. In the lower panel we show the behaviour of the universal horizon $\xi_\note{UH}$, the outer spin-0 horizon $\xi_c^+$, and the outer metric horizon $\xi_+$. As expected the three horizons tend to merge at $\xi_\note{ex}$ for small $\epsilon$, but with different scalings. With the approximate fit $N_\epsilon(\xi)\simeq 1-\xi-\epsilon^2 \xi^2(1+\xi)/2$, we get a quadratic scaling for the universal horizon $\xi_\note{UH}\simeq 1-\epsilon^2$ [by solving $N(\xi_\note{UH})=0$]. Analogously, the spin-0 horizon has a linear dependence on $\epsilon$ as $\xi_c^+\simeq1-\epsilon/s$ [Eq. \eqref{eq:spin0hor}]. Straightforwardly, for small $\epsilon$ one finds $\xi_+\simeq1-\epsilon$.

We close this section with a comment about backreaction, which we have ignored above, as we have used the decoupling limit. We can employ the approximation $N(\xi)\approx 1-\xi-\frac{\epsilon^2}{2}\xi^2(1+\xi)$ we derived earlier, which should work well for small $\epsilon$, in order to evaluate the stress-energy tensor for $u_\mu$ or $T$ field \eqref{eq:aeSEtensor}. Up to second order in $\epsilon$
\beq
T^{\aaee\,\mu}_{\;\;\;\;\;\nu}=T^{\aaee\,\mu}_{(0)\;\nu}+\epsilon^2T^{\aaee\,\mu}_{(2)\;\nu}
\eeq
where
\begin{align}
T^{\aaee\,\mu}_{(0)\;\nu}=&
\frac{c_{14} r_s^2}{8r^4}{\rm \,diag}(1,1,-1,-1)\\
T^{\aaee\,\mu}_{(2)\;\nu}=-\frac{r_s^2}{32r^6}\,{\rm \,diag}(
  &8c_{14} r^2+4c_{14} r r_s+6c_{13} r_s^2, \nonumber\\
 -&4c_{14} r_s r-3c_{14} r_s^2+6c_{13}  r_s^2, \nonumber\\
  &4c_{14} r_s r+3c_{14} r_s^2-12c_{13} r_s^2 , \nonumber\\
  &4c_{14} r_s r+3c_{14} r_s^2-12c_{13} r_s^2)
\end{align}
The resulting stress-energy tensor is manifestly parametrically small which justifies the earlier assumption of ignoring backreaction.

\begin{figure}
  \centering
  \includegraphics[width=\columnwidth]{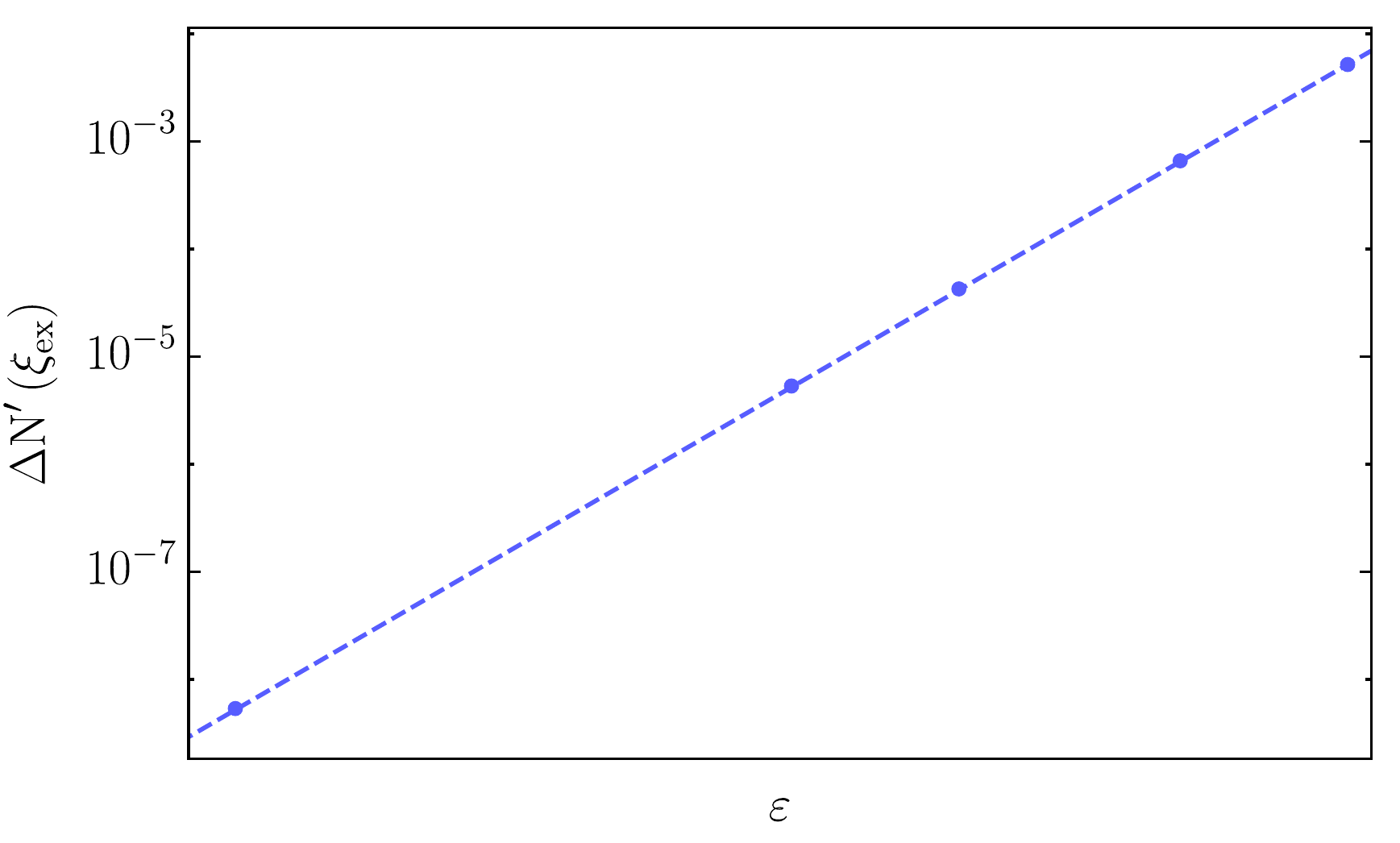}
  \caption{Mismatch of $N'$ evaluated at the matching point $\xi_\note{ex}$. The dashed line shows the $\epsilon^3$ behavior} \label{fig:nprime}
\end{figure}

\subsubsection{Interior of the universal horizon}
\label{interior}

We now want to describe the behavior of the aether inside the universal horizon. As argued before, when integrating from the outer spin-0 horizon inwards, upon reaching another spin-0 horizon, one can encounter two possibilities: (a) the solution is ``accidentally'' smooth; (b) a singularity arises on the inner spin-0 horizon.
It is not trivial to check numerically which of the above is true. If the solution were accidentally smooth, both the numerator and the denominator in the right-hand side of \eqref{eq:Nsecond} would vanish at the spin-0 horizon. This would introduce an uncontrollable error in the numerical integration which would be hard to distinguish from an actual singularity.

To overcome this difficulty, we proceed as follows. We generate a new solution, starting from the second spin-0 horizon. We select a value for $N$ and Eq.~\eqref{eq:spin0hor} is used to determine the value of $\xi$ there. Regularity uniquely determines $N'$ and then we integrate outwards, in precisely the same fashion as we did for the outermost horizon. This generates a 1-parameter family of solutions. This parameter can be thought of as the value of $N$ on the second spin-0 horizon.
By changing $N$, we then attempt to match this solution to the solution we used for the exterior at some radius. We chose to do the match at the radius of the extreme horizon (or $\xi_\note{ex}$), since this location always lies between the two spin-0 horizons.

Since Eq. \eqref{eq:Nsecond} is a second order differential equation, $N$ and $N'$ have to match at the matching point to have a smooth solution. However, we have only one parameter to tune to achieve a matching, the value on $N$ on the second spin-0 horizon. If a solution exists that is accidentally smooth on the second spin-0 horizon, then when one matches $N$ at the matching point, $N'$ should show no discontinuity as well. This is the hypothesis we have checked and it turns out not to be valid. Fig.~\ref{fig:nprime} shows the difference between the derivatives of two solutions on the matching point. It is important to mention that we have attempted to change the accuracy of the matching in $N$ or the overall accuracy of our solutions and recalculated the mismatch in $N'$. $\Delta N'$ has shown no sensitivity to such changes, so it seems sensible to conclude that there is a genuine discontinuity.

The above implies that asymptotically flat solutions with a regular outermost spin-0 horizon will exhibit a singularity on the second spin-0 horizon. Note that the decoupling limit that we are using here cannot be trusted anymore near the singularity. Within our approach this appears to be a singularity in the foliation only, because backreaction is neglected. However, one expects that, beyond decoupling, backreaction will render the spin-0 metric singular on the second spin-0 horizon.

Since the second spin-0 horizon is located inside both the outermost spin-0 and the universal horizon, this singularity poses no threat to causality --- it is not naked in any sense. However, one can be rightfully worried about the role of this singularity as $\epsilon\to 0$, as the second spin-0 horizon approaches the universal horizon and the outermost spin-0 horizon. In fact, this is precisely the limit we have studied in the previous section. Fig.~\ref{fig:nprime} clearly shows that $\Delta N'$ actually decreases as $\epsilon\to 0$. For larger value of $\epsilon$, $\Delta N'$ is well above numerical accuracy, but it scales as $\epsilon^3$. The distance between the two spin-0 horizon scales as $\epsilon$. Remarkably, for any given accuracy for numerical solutions, one can always find some sufficiently small $\epsilon$, such that $\Delta N'$ would appear to be continuous within numerical tolerance. This strongly suggests that the singularity on the second spin-0 horizon does not affect the numerical calculation that has been used to generate the exterior solution as $\epsilon \to 0$.

\section{Discussion}

We have studied black holes in the extremal limit in Einstein-aether theory and the low-energy limit of Ho\v rava gravity. We have used a decoupling limit approximation which reduces the problem to determining the preferred foliation of a Reissner-Nordstr{\"o}m spacetime. For a black hole of given mass and charge and for flat asymptotics there exists a foliation that has a universal horizon and is regular on it and everywhere in its exterior. Our numerical results strongly suggest that this foliation is unique.  However, it also appears to be singular at a specific location in the interior of the universal horizon. This location corresponds to an inner spin-0 horizon: a Killing horizon of the effective metric that defines the propagation cones of spin-0 excitations. This singularity, and the universal horizon, are always cloaked by an outermost spin-0 horizon.

The decoupling approximation that we have used here clearly breaks in the vicinity of the singular inner spin-0 horizon. However, our results strongly indicate that the decoupling approximation is valid for the exterior of the universal horizon and that our exterior solution is unaffected by the existence of this singularity, even at the extremal limit.

As expected, as extremality is approached, the universal horizon, the Killing horizons of the metric, and the outer and inner spin-0 horizons, all move toward the location of the extremal Killing horizon. The exterior of the universal horizon smoothly approaches an analytic solution in which preferred time slices are constant $t$ slices, where  $t$ is the standard time coordinate in the Reissner-Nordstr{\"o}m line element. Clearly, such a solution is not well defined on the extremal Killing horizon and on the extremal universal horizon, yet it offers a well defined description of the exterior.

Our results provide important hints for the appropriate definition of an extremal universal horizon. In every nearly-extremal solution, $u\cdot\chi$ vanishes at constant radius surface, whose radius is larger than the radius of the Killing horizon in the extremal solution. This surface can be identified as the universal horizon. It is a spacelike surface and a leaf of the preferred foliation that fails to reach spatial infinity, well in line with the definitions and results of Ref.~\cite{Bhattacharyya:2015gwa}. Moreover, $a\cdot\chi$, which can be thought of as analogous to surface gravity \cite{Berglund:2012fk,Cropp:2013sea,Bhattacharyya:2015gwa}, does not vanish on that surface for any non-extremal solution.

It is important to examine the behaviour of the foliation in the extremal limit near the extremal Killing horizon: as  $Q^2/M^2 \to 1$,  $a\cdot\chi\to 0$ everywhere in the exterior and $u\cdot\chi\to 0$ as $r\to r_{\rm ex}$. That is, all leaves that span the exterior appear to asymptote and ``pile up'' on the Killing horizon at the extremal limit. Remarkably, the extremal Killing horizon appears to satisfy the requirements
of the global definition of a universal horizon, as given in Ref.~\cite{Bhattacharyya:2015gwa}. The fact that $a\cdot\chi\to 0$ in the extremal limit suggests that such a universal horizon is degenerate in the appropriate sense, as expected. However, most of the rest of the definitions and proofs of Ref.~\cite{Bhattacharyya:2015gwa} are inapplicable to the extremal solution. A key assumption of Ref.~\cite{Bhattacharyya:2015gwa} is that every point of the manifold belongs to a leaf of the foliation and our analysis indicates that points that lie on the extremal Killing horizon cannot satisfy this property.  Indeed, a Killing horizon cannot be a leaf of the foliation, as it is a null surface.  Moreover, when $a\cdot\chi= 0$, the proof of Ref.~\cite{Bhattacharyya:2015gwa} that a universal horizon must be a leaf (Proposition 2 in Sec. 4) does not apply.\footnote{Note that Theorem 4 in the same section relies heavily of Proposition 2.}

It would be interesting to study nearly extremal and extremal black holes with universal horizons without using a decoupling approximation. It is also important to consider spinning as opposed to electrically charged black holes. It is also important to understand the role of extremal universal horizons in the context of universal horizon thermodynamics (see {\em e.g.}~\cite{Berglund:2012bu,Berglund:2012fk,Cropp:2013sea,Michel:2015rsa,Pacilio:2017emh}.
\begin{acknowledgments}
The authors thank Sergey Sibiryakov for helpful discussions.
The research leading to these results has received funding  from the European Research Council under the European Union Seventh Framework Programme (FP7/2007-2013) / ERC Grant Agreement No.~306425 ``Challenging General Relativity.'' M.S. is supported by the Royal Commission for the Exhibition of 1851.
\end{acknowledgments}

\bibliography{bibnote}

\end{document}